**Spin-orbit torque in MgO/CoFeB/Ta/CoFeB/MgO symmetric structure with interlayer antiferromagnetic coupling**

G. Y. Shi,[1] C. H. Wan,[2] Y. S. Chang,[2] F. Li,[1] X. J. Zhou,[1] P. X. Zhang,[1] J. W. Cai,[2,*] X. F. Han,[2] F. Pan,[1] and C. Song[1,†]

[1] Key Laboratory of Advanced Materials (MOE), School of Materials Science and Engineering, Tsinghua University, Beijing 100084, China

[2] Beijing National Laboratory for Condensed Matter Physics, Institute of Physics, Chinese Academy of Sciences, Beijing 100190, China

Spin current generated by spin Hall effect in the heavy metal would diffuse up and down to adjacent ferromagnetic layers and exert torque on their magnetization, called spin-orbit torque. Antiferromagnetically coupled trilayers, namely the so-called synthetic antiferromagnets (SAF), are usually employed to serve as the pinned layer of spintronic devices based on spin valves and magnetic tunnel junctions to reduce the stray field and/or increase the pinning field. Here we investigate the spin-orbit torque in MgO/CoFeB/Ta/CoFeB/MgO perpendicularly magnetized multilayer with interlayer antiferromagnetic coupling. It is found that the magnetization of two CoFeB layers can be switched between two antiparallel states simultaneously. This observation is replicated by the theoretical calculations by solving Stoner-Wohlfarth model and Landau-Lifshitz-Gilbert equation. Our findings combine spin-orbit torque and interlayer coupling, which might advance the magnetic memories with low stray field and low power consumption.

* jwcai@iop.ac.cn
† songcheng@mail.tsinghua.edu.cn

The spin Hall effect (SHE), a robust way to generate spin current, is a transport phenomenon which demonstrates that an electric current flows through nonmagnetic materials and generates orthogonal spin polarization and spin current [1–5]. The efficiency of the charge to spin conversion, characterized by spin Hall angle, generally depends on the strength of spin-orbit coupling [6–8]. For the in-plane case, up- and down-polarized spins accumulate at the edge of the channel, which was directly detected via Kerr microscope [3] and Hanle effect [5]. For the out-of-plane case, spins with opposite directions diffuse upward and downward, which exert torques to neighboring magnetic layers. This is so-called spin-orbit torque (SOT), which is considered as an effective way to switch magnetization [9–19] and to drive domain wall motion [20,21] with low power consumption. The torque includes two components, e.g., damping-like torque and field-like torque, which have different symmetries with respect to magnetization reversal [22–24]. These torques essentially generated by spin-orbit coupling have also been demonstrated both theoretically and experimentally in antiferromagnetic systems. Gomonay *et al*. [25,26] proposed that damping-like torque induced by spin-polarized current can produce large angle reorientation of antiferromagnetic magnetization. And Wadley *et al*. [27] showed the electrical switching of antiferromagnetic CuMnAs via field-like torque generated by non-equilibrium spin polarizations. In general, such a system which consists of a heavy metal (HM)/ferromagnetic metal (FM)/oxide heterostructure only utilizes one side of the spin current brought by SHE. Moreover, Woo *et al*. [28] constructed a Pt/Co/Ta structure to enhance spin-orbit torque due to the opposite sign of spin Hall angle of Pt and Ta, whereas even in this structure only one side of the spin polarization produced by the SHE of the heavy metals is available for the magnetization switching. Obviously, there is a pressing need to develop a different structure, e.g., CoFeB/Ta/CoFeB sandwich, to make use of both spin currents from the Ta layer flowing upward and downward, to realize the magnetization switching of two adjacent CoFeB layers.

The interlayer coupling in ferromagnetic/non-magnetic/ferromagnetic sandwich structures has been extensively studied. It is generally accepted that the coupling oscillates between ferromagnetic and antiferromagnetic as the non-magnetic inset layer (such as Ru and Cr) thickness increases with a long oscillation period of ~1 nm, ascribed to Ruderman-Kittel-Kasuya-Yosida (RKKY) exchange interaction [29–31]. However, the oscillation vanishes and only antiferromagnetic coupling remains in the sandwich structures with bcc heavy metals, e.g., Nb, Ta, and W [30]. Thus, antiferromagnetic interlayer coupling is expected to play an important role in the SOT-induced magnetization switching in CoFeB/Ta/CoFeB-based heterostructures. The theories and experiments below demonstrate that in perpendicularly magnetized MgO/CoFeB/Ta/CoFeB/MgO heterostructures the two CoFeB layers with antiferromagnetic coupling can be switched between two anti-parallel states simultaneously through the SOT.

MgO(4)/Co$_{40}$Fe$_{40}$B$_{20}$(1.3)/Ta(1.2)/Co$_{40}$Fe$_{40}$B$_{20}$(1.05)/MgO(2)/SiO$_2$(2) (from the bottom to the top, thickness in nanometer) heterostructures were deposited on thermally oxidized Si substrates via magnetron sputtering at a base vacuum of $5 \times 10^{-5}$ Pa. In order to optimize the perpendicular magnetic anisotropy (PMA), the films were annealed at 300 ºC for half an hour at the vacuum. After that, typical Hall bar devices with a channel width of 3 μm were fabricated by lithography and Ar ion etching. For theoretical calculations, the Stoner-Wohlfarth model [32], torque balance equation and the Landau-Lifshitz-Gilbert (LLG) equation [10] were adopted to simulate the current-induced magnetization switching in the heterostructures with PMA and antiferromagnetic coupling.

Figure 1(a) displays a schematic design of the sample layout and measurement configuration. The expanded area exhibits the concept that with the current applied to the Ta layer, the SHE creates inverse spin polarization, which diffuses up and down into the adjacent CoFeB layers, giving rise to magnetization switching. We first show in Fig. 1(b) an anomalous Hall effect (AHE) curve measured with a current ($I$) of 0.1 mA applied to the Hall bar along $+y$ direction and an external magnetic field ($H_{ext}$) along $z$ direction ($\beta = 90°$). There are two striking features for the AHE curve: i) separate switching fields for the upper and lower CoFeB layers, i.e., 72 Oe and –110 Oe for the descend branch with a plateau in between, indicating these two CoFeB layers are antiferromagnetic coupled with each other. A similar interlayer antiferromagnetic coupling is observed in a series of CoFeB/Ta($t$)/CoFeB ($t$ = 1.0, 1.2, 1.4, and 1.6 nm) without RKKY oscillation, which is consistent to the observation in Co/Ta multilayers [30]; iii) the square shape of the AHE curve confirms the PMA of the CoFeB layers, which benefits the current-induced magnetization switching via the SOT. Note that it is the $z$ component of the external field leading to the magnetization reversal when $H_{ext}$ is swept in $yz$ plane and close to $y$ axis ($\beta = 1°$), the sudden change of Hall resistance ($R_H$) occurs at a much larger external field compared to the case of $\beta = 90°$, as presented in Fig. 1(c). When $H_{ext}$ is up to more than 2 kOe, the magnetization of CoFeB is gradually aligned to near in-plane position. As a consequence, the $z$ component of the total magnetization continuously reduces, causing the decrease of the Hall resistance.

We then focus on the current-induced magnetization switching via the SOT. For these measurements, a constant external field was applied along $y$ direction and the Hall resistance was recorded while sweeping the current. The most eminent feature in Fig. 2(a) is that the magnetization switching induced by the current shows a hysteresis window with a critical current of ~2.6 mA ($J_e = 2.44 \times 10^7$ A/cm$^2$) and the switching is anticlockwise for positive $H_{ext}$ (+500 Oe) and clockwise for negative $H_{ext}$ (–500 Oe). The switching direction is similar to that of typical Ta/CoFeB/MgO structures [9]. Moreover, the quantity of the Hall resistance at two stable states (±2.5 Ω) clarifies that the magnetization switching occurs between two antiparallel states of CoFeB moments, which is consistent with the plateau resistance in the AHE curve shown in Fig. 1(b).

A comparison of the current-induced magnetization switching at various external magnetic fields is depicted in Fig. 2(b). Apparently, the critical current for magnetization switching drops with enhancing $H_{ext}$ from 500 to 1250 Oe. Also visible is the gradual decrease of the Hall resistance when the applied current is higher than the critical current. This tendency indicates that the perpendicular magnetized CoFeB would be switched to a position in the vicinity of $yz$ plane by the strong applied current. Furthermore, when the current is near 4 mA, the Hall resistances suddenly increase due to Joule heating. Particularly, as demonstrated in Fig. 2(b), the current-induced switching curves show the opposite nonlinear behavior at positive and negative value of Hall resistance before the switching. This observation is different from the previous reports in a single ferromagnetic layer system [10,14].

Now, we turn to the simulation part in order to interpret the experimental results. Before utilizing the torque balance equation to simulate current-induced magnetization switching, both the interlayer antiferromagnetic coupling and PMA features need to be involved to the magnetic state of CoFeB/Ta/CoFeB via the Stoner-Wohlfarth model. On the basis of the Stoner-Wohlfarth model, a simple structure is set up, where a heavy metal is sandwiched by two single domain ferromagnetic layers ($FM_1$ and $FM_2$) with easy-axis along $z$ axis, as illustrated in Fig. 3(a). While sweeping the external field along $z$ axis, the hysteresis loop can be obtained by solving the local minimum of free energy of this structure. The free energy consists of Zeeman energy, anisotropy energy and antiferromagnetic coupling energy,

$$E = -(H_z M_1 \sin\omega_1 + H_z M_2 \sin\omega_2) + K_1 \cos^2\omega_1 + K_2 \cos^2\omega_2 + A_{12} \cos(\omega_1 - \omega_2), \quad (1)$$

where $M_1$ and $M_2$ are the magnetization of $FM_1$ and $FM_2$, $K_1$ and $K_2$ characterize their anisotropy energy, antiferromagnetic coupling energy is expressed by $A_{12}$ and the angle between $y$ axis and the magnetization of $FM_1$ and $FM_2$ are $\omega_1$ and $\omega_2$ respectively [Fig. 3(a)]. Considering that the free energy $E$ is a function of two variables $\omega_1$ and $\omega_2$, the local minimum of free energy fulfills two conditions: One is

$$\frac{\partial E}{\partial \omega_1} = 0 \quad \text{and} \quad \frac{\partial E}{\partial \omega_2} = 0, \quad (2)$$

the other is

$$\frac{\partial^2 E}{\partial \omega_1^2} > 0 \quad \text{and} \quad \det\begin{bmatrix} \dfrac{\partial^2 E}{\partial \omega_1^2} & \dfrac{\partial^2 E}{\partial \omega_1 \partial \omega_2} \\ \dfrac{\partial^2 E}{\partial \omega_2 \partial \omega_1} & \dfrac{\partial^2 E}{\partial \omega_2^2} \end{bmatrix} > 0. \quad (3)$$

Accordingly, the hysteresis loop of the proposed structure could be obtained via solving the specific $\omega_1$ and $\omega_2$ that satisfy Eqs. (2) and (3) for each given external field $H_z$ and bringing them into $M_z = (M_1 \sin\omega_1 + M_2 \sin\omega_2)/(M_1 + M_2)$, which expresses the normalized projection of total magnetization on $z$ axis. Figure 3(b) shows a representative hysteresis loop by plugging these parameters to Eq. (1): $K_1 = 1.3 \times 10^6$ erg/cm$^3$, $K_2 = 1 \times 10^6$ erg/cm$^3$, $A_{12} = 2 \times 10^6$ erg/cm$^3$, $M_1 = 1500$ emu/cm$^3$ and $M_2 = 1300$ emu/cm$^3$. Remarkably, the shape of the hysteresis loop reflects both the PMA and antiferromagnetic coupling in the proposed structure.

The static evolution of the magnetization of $FM_1$ and $FM_2$ layers can be derived by the torque balance equation [10],

$$\begin{aligned}\tau_{tot1} = \tau_{ext1} + \tau_{an1} + \tau_{coup1} - \tau = 0 \\ \tau_{tot2} = \tau_{ext2} + \tau_{an2} + \tau_{coup2} + \tau = 0\end{aligned}, \quad (4)$$

where the torques on the magnetic moment including external field torque $\tau_{ext}$, anisotropy field torque $\tau_{an}$, antiferromagnetic coupling field torque $\tau_{coup}$ and spin-orbit torque $\tau$. It is worthy pointing out that in the torque balance equation of $FM_1$ the sign of $\tau$ is negative due to the negative spin Hall angle of HM assumed in the proposed model. While in the torque balance equation of $FM_2$, the sign of $\tau$ is positive considering that the spin polarization induced by SHE is opposite for spins moving to two opposite directions. For clarity, the scalar expression of torque balance equation for $FM_1$ and $FM_2$ is derived from the vector form in Eq. (4) [see Appendix I for detailed derivation]. The magnetic parameters adopted in Eq. (4) are identical to that of Eq. (1) and the corresponding results are presented in Fig. 3(c) and (d). The magnetic moment $\mathbf{m_1}$ and $\mathbf{m_2}$ can be rotated in $yz$ plane and switched simultaneously for a certain amount of $\tau$.

As Fig. 3(c) shows, the $z$ components of $\mathbf{m_1}$ and $\mathbf{m_2}$ keep opposite due to the antiferromagnetic coupling in the whole process. Moreover, for a positive external field (e.g., $H_y$ = 1000 Oe) and a positive current, $\mathbf{m_1}$ prefer to point up and $\mathbf{m_2}$ prefer to point down. Differently, for a positive external field and a negative current, $\mathbf{m_1}$ tend to point down while $\mathbf{m_2}$ does the opposite. This scenario reveals that the switching is anticlockwise for positive $H_y$ considering that $M_1$ is stronger than $M_2$ in the present structure as Fig. 3(d) depicts. The situation differs dramatically when a negative external field (e.g., $H_y$ = −1000 Oe) is used, the current-induced magnetization switching is clockwise. This indicates that the switching symmetry is in consistent with the experimental results shown in Fig 2(a). Moreover, as Fig. 3(d) shows, $M_z$ exhibits opposite nonlinear behavior before switching. Thus, the opposite nonlinear behavior of Hall resistance before switching in Fig. 2(b) is well replicated by simulation and can be ascribed to the opposite switching of the upper and lower CoFeB and their combination. What's more, as expected, with the increase of $H_y$ from 250 to 1500 Oe, critical $\tau$ for magnetization switching is reduced greatly, as displayed in Fig. 3(e), indicating that the critical current density is inversely proportionally to the external field within this range of $H_y$.

The dynamic evolution of the magnetic moment of the present CoFeB/Ta/CoFeB structure with PMA and antiferromagnetic coupling can be described by performing macrospin simulation on the basis of LLG equation,

$$\frac{d\hat{\mathbf{m}}}{dt} = -\gamma \hat{\mathbf{m}} \times \mathbf{H}_{\text{eff}} + \alpha \hat{\mathbf{m}} \times \frac{d\hat{\mathbf{m}}}{dt} + \gamma \zeta_{\parallel} \hat{\mathbf{m}} \times (\hat{\mathbf{m}} \times \hat{\boldsymbol{\sigma}}) + \gamma \zeta_{\perp} \hat{\mathbf{m}} \times \hat{\boldsymbol{\sigma}}, \quad (5)$$

where $\hat{\mathbf{m}}$ represents the unit magnetization moment vector and its orientation is defined in spherical coordinates as depicted in Fig. 4(a), $\hat{\boldsymbol{\sigma}}$ is the spin polarization collinear to $x$ axis given that we assume the current is along $y$ direction, $\gamma$ is gyromagnetic ratio and $\alpha$ is damping constant. The effective field $\mathbf{H}_{\text{eff}}$, which has two orthogonal components along polar angle direction $H_\theta$ and azimuth angle direction $H_\varphi$, is composed of external field, anisotropy field and antiferromagnetic coupling effective field. It can be derived from the free energy of our system [see Appendix II]. Damping-like torque coefficient is described by $\zeta_{\parallel} = \frac{\hbar c_{\parallel} J_e}{2eM_s t_F}$, where $c_{\parallel}$ is damping-like torque efficiency, $J_e$ is current density, $M_s$ is saturation magnetization per unit volume and $t_F$ denotes the thickness of the ferromagnetic layer. $\zeta_{\perp} = \frac{\hbar c_{\perp} J_e}{2eM_s t_F}$ is the field-like torque coefficient and field-like torque efficiency is represented by $c_\perp$. To make the LLG equation more convenient for calculation, $\mathbf{H}_{eff}$ is normalized by the anisotropy effective field of FM$_2$, $H_{an2}$ [see Appendix II for detailed derivation process]. Therefore, the LLG equations for the upper FM$_1$ and lower FM$_2$ take the dimensionless form of

$$\frac{d\hat{\mathbf{m}}_1}{dt} = -\hat{\mathbf{m}}_1 \times \frac{\mathbf{h}_{eff1}}{g} + \alpha \hat{\mathbf{m}}_1 \times \frac{d\hat{\mathbf{m}}_1}{dt} + C_\parallel \frac{1}{g} \hat{\mathbf{m}}_1 \times (\hat{\mathbf{m}}_1 \times \hat{\boldsymbol{\sigma}}) + C_\perp \frac{1}{g} \hat{\mathbf{m}}_1 \times \hat{\boldsymbol{\sigma}}$$
$$\frac{d\hat{\mathbf{m}}_2}{dt} = -\hat{\mathbf{m}}_2 \times \frac{\mathbf{h}_{eff2}}{g} + \alpha \hat{\mathbf{m}}_2 \times \frac{d\hat{\mathbf{m}}_2}{dt} - C_\parallel \frac{1}{g} \hat{\mathbf{m}}_2 \times (\hat{\mathbf{m}}_2 \times \hat{\boldsymbol{\sigma}}) - C_\parallel \frac{1}{g} \hat{\mathbf{m}}_2 \times \hat{\boldsymbol{\sigma}} \quad , (6)$$

where $\frac{1}{g} = \frac{\gamma H_{an2}}{2}$ and normalized torque coefficient $C_\parallel = \gamma \zeta_\parallel g = \frac{\hbar c_\parallel J_e}{eM_s t_F H_{an2}}$ and $C_\perp = \gamma \zeta_\perp g = \frac{\hbar c_\perp J_e}{eM_s t_F H_{an2}}$. For simplicity, the upper FM$_1$ layer is supposed to possess the same anisotropy and saturation magnetization as the lower FM$_2$ layer, which do not influence the main results of this simulation. Firstly, the initial position of $m_1$ and $m_2$ are set to ensure that they are nearly antiparallel and have a little tilt angle off $z$ axis. Parameters $M_s$ = 1300 emu/cm$^3$, $t_F$ = 10$^{-7}$ cm, $H_{an2}$ = 1333.33 Oe and $\alpha$ = 0.01 were brought into Eq. (6). As a result, magnetization switching trajectories under three typical values of normalized torque coefficient [$C_\parallel = 0.75$ $C_\perp = 0.4$, $C_\parallel = 1.5$ $C_\perp = 0.8$, and $C_\parallel = 3$ $C_\perp = 1.6$ for Fig. 4(b), 4(c), and 4(d), respectively] were calculated with fixed assistant external field and antiferromagnetic coupling, corresponding to three typical quantity of current density. As depicted in Fig. 4(b), when the damping-like torque and field-like torque coefficients are 0.75 and 0.4 respectively, the magnetic moments $m_1$ and $m_2$ precess around the final positions near the initial positions, indicating that magnetization switching does not take place under this torque coefficient value. With the damping-like torque and field-like torque coefficient separately increasing up to 1.5 and 0.8, the magnetic moments $m_1$ and $m_2$ quickly rotate to the opposite hemisphere and then precess around the final equilibrium positions, as shown in Fig. 4(c). Figure 4(d) shows that when $C_{\|}$ and $C_{\perp}$ rise up to 3 and 1.6, both magnetic moments switch to the opposite hemisphere and stay at the stable position without apparent precession.

We then turn towards the time dependent magnetic moments projection on $z$ axis. Corresponding data are presented in Fig. 4(e). For $C_{\|} = 0.75$ and $C_{\perp} = 0.4$, $m_1$ and $m_2$ move around the equilibrium positions near the initial positions without magnetization switching. For $C_{\|} = 1.5$ and $C_{\perp} = 0.8$, the switching of $m_1$ and $m_2$ happens simultaneously, which is less than three nanoseconds, accompanied by a relatively long precession around the final states. When $C_{\|}$ is up to 3 and $C_{\perp}$ is up to 1.6, $m_1$ and $m_2$ rapidly switch up and down to the final states. It is worthy pointing out that the $z$ components of $m_1$ and $m_2$ exhibit the same amount about 0.31 but with opposite sign. This indicates that if the torque coefficient is large enough, the $z$ components of $m_1$ and $m_2$ would decrease. As a consequence, the Hall resistance would reduce at a certain amount of applied current density, which is observed in Fig. 2.

In conclusion, through the spin-orbit torque experiments in MgO/CoFeB/Ta/CoFeB/MgO symmetric heterostructures with PMA and antiferromagnetic coupling, we demonstrate that the spin current generated by spin Hall effect of Ta diffuse up and down to adjacent CoFeB layers and the magnetization of two CoFeB layers can be switched between two antiparallel states with a critical current density of ~$10^7$ A/cm$^2$. The experimental results can be well reproduced by simulation. Our findings on spin-orbit torque in the antiferromagnetic coupling system might advance the magnetic memories with low stray field and low power consumption [33].

This work was supported by the National Natural Science Foundation of China (Grant Nos. 51571128, 51671110, and 51371191) and Ministry of Science and Technology of the People's Republic of China (2016YFA0203800).

**Appendix I: Derivation of torque balance equations**

The torque balance equation means that the total torques exerted on magnetic moment are equal to zero, which includes the external field torque, perpendicular anisotropy field torque, antiferromagnetic coupling field torque and spin torque. Then, for the upper ferromagnetic layer $FM_1$ and lower ferromagnetic layer $FM_2$, torque balance equations are expressed as,

$$\begin{aligned}\boldsymbol{\tau}_{tot1} &= -\mathbf{M_1} \times \mathbf{H}_{eff1} - \boldsymbol{\tau} = \boldsymbol{\tau}_{ext1} + \boldsymbol{\tau}_{an1} + \boldsymbol{\tau}_{coup1} - \boldsymbol{\tau} = 0 \\ \boldsymbol{\tau}_{tot2} &= -\mathbf{M_2} \times \mathbf{H}_{eff2} + \boldsymbol{\tau} = \boldsymbol{\tau}_{ext2} + \boldsymbol{\tau}_{an2} + \boldsymbol{\tau}_{coup2} + \boldsymbol{\tau} = 0\end{aligned}, \text{(A.1)}$$

With an external field being fixed on $y$ axis, the free energy of the system is given by

$$E = -(H_y M_1 \cos\omega_1 + H_y M_2 \cos\omega_2) + K_1 \cos^2\omega_1 + K_2 \cos^2\omega_2 + A_{12}\cos(\omega_1 - \omega_2), \text{(A.2)}$$

and the effective field for $M_1$ and $M_2$ is expressed as

$$\begin{aligned}\mathbf{H}_{eff1} &= -\mathbf{e}_\omega \frac{\partial E}{M_1 \partial \omega_1} = -\mathbf{e}_\omega (H_y \sin\omega_1 - 2\frac{K_1}{M_1}\cos\omega_1 \sin\omega_1 - \frac{A_{12}}{M_1}\sin(\omega_1 - \omega_2)) \\ \mathbf{H}_{eff2} &= -\mathbf{e}_\omega \frac{\partial E}{M_1 \partial \omega_2} = -\mathbf{e}_\omega (H_y \sin\omega_2 - 2\frac{K_2}{M_2}\cos\omega_2 \sin\omega_2 + \frac{A_{12}}{M_2}\sin(\omega_1 - \omega_2))\end{aligned}. \text{(A.3)}$$

If $\tau$ is not large enough, $\mathbf{m_1}$ and $\mathbf{m_2}$ can be proven to remain in $yz$ plane. Under this situation, all torques lie in $x$ axis and the torque balance equations take a simple form of

$$\begin{aligned}\tau_{tot1} &= \mathbf{e_x} \cdot \boldsymbol{\tau}_{tot1} = -M_1 H_{eff1} - \tau = H_y M_1 \sin\omega_1 - 2K_1 \cos\omega_1 \sin\omega_1 - A_{12}\sin(\omega_1 - \omega_2) - \tau = 0 \\ \tau_{tot2} &= \mathbf{e_x} \cdot \boldsymbol{\tau}_{tot2} = -M_2 H_{eff2} - \tau = H_y M_2 \sin\omega_2 - 2K_2 \cos\omega_2 \sin\omega_2 + A_{12}\sin(\omega_1 - \omega_2) + \tau = 0\end{aligned}. \text{(A.4)}$$

**Appendix II: Derivation of effective field in LLG equation**

In spherical coordinate, the free energy of our system is expressed as,

$$E = -H_y M_1 \sin\theta_1 \sin\varphi_1 - H_y M_2 \sin\theta_2 \sin\varphi_2 + K_1 \sin^2\theta_1 + K_2 \sin^2\theta_2 + A_{12}\cos\langle\mathbf{m_1},\mathbf{m_2}\rangle, \text{(A.5)}$$

where $\cos\langle\mathbf{m_1},\mathbf{m_2}\rangle$ means the cosine of the included angle of $\mathbf{m_1}$ and $\mathbf{m_2}$ which can be derived by cosine law and writes

$$\cos\langle \mathbf{m_1}, \mathbf{m_2}\rangle = \sin\theta_1 \cos\varphi_1 \sin\theta_2 \cos\varphi_2 + \sin\theta_1 \sin\varphi_1 \sin\theta_2 \sin\varphi_2 + \cos\theta_1 \cos\theta_2. \quad (A.6)$$

Thus, $\mathbf{H}_{eff}$ for $FM_1$ and $FM_2$ are expressed as

$$\begin{aligned}
\mathbf{H}_{eff1} &= \mathbf{e}_\theta H_{\theta_1} + \mathbf{e}_\varphi H_{\varphi_1} \\
H_{\theta_1} &= -\frac{\partial E}{M_1 \partial \theta_1} = -\frac{1}{M_1}(-H_y M_1 \cos\theta_1 \sin\varphi_1 + K_1 \sin 2\theta_1 + A_{12}\cos\theta_1 \cos\varphi_1 \sin\theta_2 \cos\varphi_2 + A_{12}\cos\theta_1 \sin\varphi_1 \sin\theta_2 \sin\varphi_2 \\
&\quad - A_{12}\sin\theta_1 \cos\theta_2) \\
H_{\varphi_1} &= -\frac{\partial E}{M_1 \sin\theta_1 \partial\varphi_1} = -\frac{1}{M_1}(-H_y M_1 \cos\varphi_1 - A_{12}\sin\varphi_1 \sin\theta_2 \cos\varphi_2 + A_{12}\cos\varphi_1 \sin\theta_2 \sin\varphi_2) \\
\mathbf{H}_{eff2} &= \mathbf{e}_\theta H_{\theta_2} + \mathbf{e}_\varphi H_{\varphi_2} \\
H_{\theta_2} &= -\frac{\partial E}{M_2 \partial \theta_2} = -\frac{1}{M_2}(-H_y M_2 \cos\theta_2 \sin\varphi_2 + K_2 \sin 2\theta_2 + A_{12}\sin\theta_1 \cos\varphi_1 \cos\theta_2 \cos\varphi_2 + A_{12}\sin\theta_1 \sin\varphi_1 \cos\theta_2 \sin\varphi_2 \\
&\quad - A_{12}\cos\theta_1 \sin\theta_2) \\
H_{\varphi_2} &= -\frac{\partial E}{M_2 \sin\theta_2 \partial\varphi_2} = -\frac{1}{M_2}(-H_y M_2 \cos\varphi_2 - A_{12}\sin\theta_1 \cos\varphi_1 \sin\varphi_2 + A_{12}\sin\theta_1 \sin\varphi_1 \cos\varphi_2)
\end{aligned}$$

(A.7)

Taking the different magnetic parameters of two FM layers into account, we normalize $\mathbf{H}_{eff}$ by the anisotropy effective field of $FM_2$ $H_{an2}$. Hence,

$$\begin{aligned}
\mathbf{h}_{eff1} &= \mathbf{e}_\theta h_{\theta_1} + \mathbf{e}_\varphi h_{\varphi_1} \\
h_{\theta_1} &= \frac{2H_{\theta_1}}{H_{an2}} = -\frac{1}{m_{12}}(-2h_y m_{12}\cos\theta_1 \sin\varphi_1 + k\sin 2\theta_1 + a\cos\theta_1 \cos\varphi_1 \sin\theta_2 \cos\varphi_2 + a\cos\theta_1 \sin\varphi_1 \sin\theta_2 \sin\varphi_2 - a\sin\theta_1 \cos\theta_2) \\
h_{\varphi_1} &= \frac{2H_{\varphi_1}}{H_{an2}} = -\frac{1}{m_{12}}(-2h_y m_{12}\cos\varphi_1 - a\sin\varphi_1 \sin\theta_2 \cos\varphi_2 + a\cos\varphi_1 \sin\theta_2 \sin\varphi_2) \\
\mathbf{h}_{eff2} &= \mathbf{e}_\theta h_{\theta_2} + \mathbf{e}_\varphi h_{\varphi_2} \\
h_{\theta_2} &= \frac{2H_{\theta_2}}{H_{an2}} = 2h_y \cos\theta_2 \sin\varphi_2 - \sin 2\theta_2 - a\sin\theta_1 \cos\varphi_1 \cos\theta_2 \cos\varphi_2 - a\sin\theta_1 \sin\varphi_1 \cos\theta_2 \sin\varphi_2 + a\cos\theta_1 \sin\theta_2 \\
h_{\varphi_2} &= \frac{2H_{\varphi_2}}{H_{an2}} = 2h_y \cos\varphi_2 + a\sin\theta_1 \cos\varphi_1 \sin\varphi_2 - a\sin\theta_1 \sin\varphi_1 \cos\varphi_2
\end{aligned}$$

(A.8)

Where $h_y = H_y / H_{an2}$, $a = \frac{A_{12}}{K_2}(K_2 = \frac{1}{2}H_{an2}M_2)$, $k = \frac{K_1}{K_2}$ and $m_{12} = \frac{M_1}{M_2}$. Therefore, the LLG equations for upper $FM_1$ and lower $FM_2$ take the form of,

$$\begin{aligned}
\frac{d\hat{\mathbf{m}}_1}{dt} &= -\hat{\mathbf{m}}_1 \times \frac{\mathbf{h}_{eff1}}{g} + \alpha\hat{\mathbf{m}}_1 \times \frac{d\hat{\mathbf{m}}_1}{dt} + C_\parallel \frac{1}{g}\hat{\mathbf{m}}_1 \times (\hat{\mathbf{m}}_1 \times \hat{\boldsymbol{\sigma}}) + C_\perp \frac{1}{g}\hat{\mathbf{m}}_1 \times \hat{\boldsymbol{\sigma}} \\
\frac{d\hat{\mathbf{m}}_2}{dt} &= -\hat{\mathbf{m}}_2 \times \frac{\mathbf{h}_{eff2}}{g} + \alpha\hat{\mathbf{m}}_2 \times \frac{d\hat{\mathbf{m}}_2}{dt} - C_\parallel \frac{1}{g}\hat{\mathbf{m}}_2 \times (\hat{\mathbf{m}}_2 \times \hat{\boldsymbol{\sigma}}) - C_\parallel \frac{1}{g}\hat{\mathbf{m}}_2 \times \hat{\boldsymbol{\sigma}}
\end{aligned}, \quad (A.9)$$

where $\frac{1}{g} = \frac{\gamma H_{an2}}{2}$ and normalized damping-like torque coefficient

$$C_\parallel = \gamma \zeta_\parallel g = \frac{\hbar c_\parallel J_e}{eM_s t_F H_{an2}} \quad \text{and field-like torque coefficient} \quad C_\perp = \gamma \zeta_\perp g = \frac{\hbar c_\perp J_e}{eM_s t_F H_{an2}}.$$

Figure Captions

FIG. 1. (color online) (a) Schematic of MgO/CoFeB/Ta/CoFeB/MgO multilayer. Expand area shows SHE brought on spin polarization and in turn magnetization switching. $R_H$ curves measured when (b) $H_{ext}$ is fixed in +z direction and (c) $H_{ext}$ is in the yz plane and near +y direction ($\beta = 1°$).

FIG. 2. (color online) (a) Magnetization switching characterized by $R_H$ in the presence of a positive and negative external field fixed at y direction. (b) Current-induced switching under different external magnetic fields applied in +y direction.

FIG. 3. (color online) (a) Sketch of a Heavy metal layer sandwiched by two ferromagnetic metal layers and corresponding Cartesian coordinates with relevant orientation of magnetic moment $\mathbf{m_1}$ and $\mathbf{m_2}$. (b) Hysteresis loop calculated by Stoner-Wohlfarth model. Current-induced switching under positive and negative external field exhibited by the angular coordinate of $\mathbf{m_1}$ and $\mathbf{m_2}$ (c) and z component of total magnetization $M_z$ (d). (e) Critical $\tau$ for magnetization switching vs external field by solving torque balance equation.

FIG. 4. (color online) (a) Orientation of magnetic moment $\mathbf{m_1}$ and $\mathbf{m_2}$ defined in macrospin calculation. Magnetization switching trajectories with the same assistant external field $h_y = 0.375$, antiferromagnetic coupling $a = 2$, and different dimensionless torque coefficients (b) $C_\parallel = 0.75$ $C_\perp = 0.4$, (c) $C_\parallel = 1.5$ $C_\perp = 0.8$, and (d) $C_\parallel = 3$ $C_\perp = 1.6$. The pink solid line and the green solid line represent the initial and final position of magnetization respectively. The red curves and the blue curves individually stand for the track of magnetic moment $\mathbf{m_1}$ and $\mathbf{m_2}$. e) Time evolution of the z component of magnetization extracted from (b), (c) and (d).

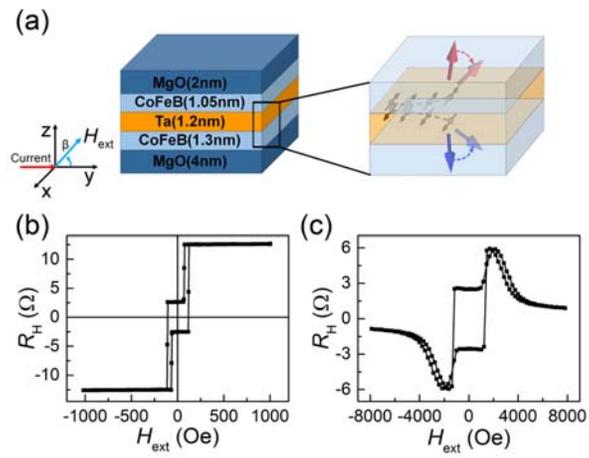

Shi *et al.*, FIG. 1.

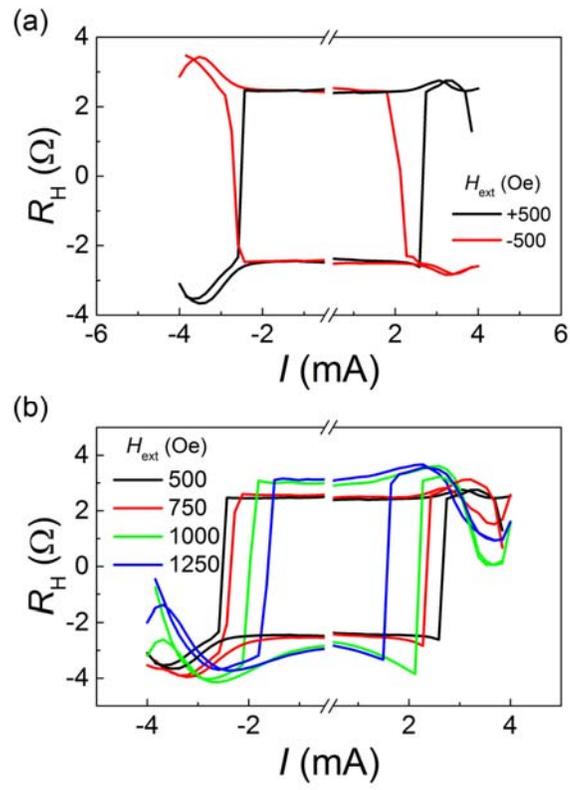

Shi *et al.*, FIG. 2.

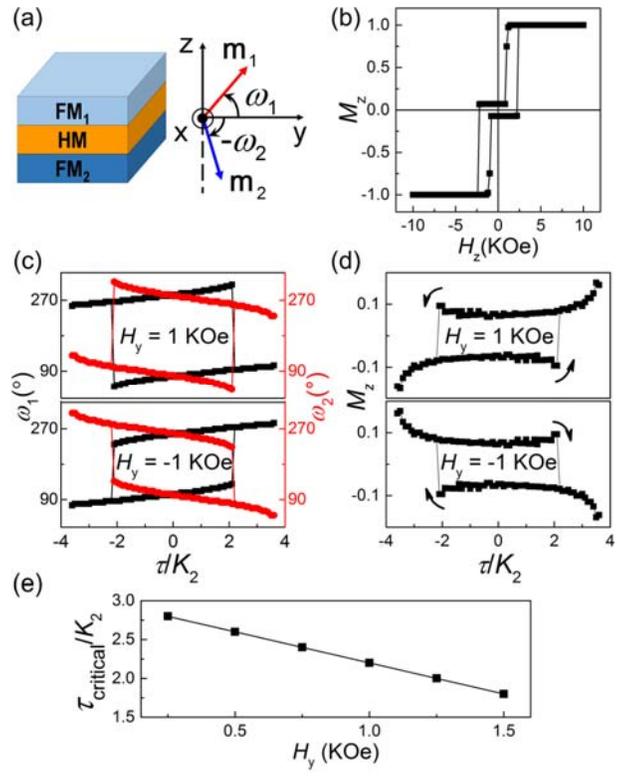

Shi *et al.*, FIG. 3.

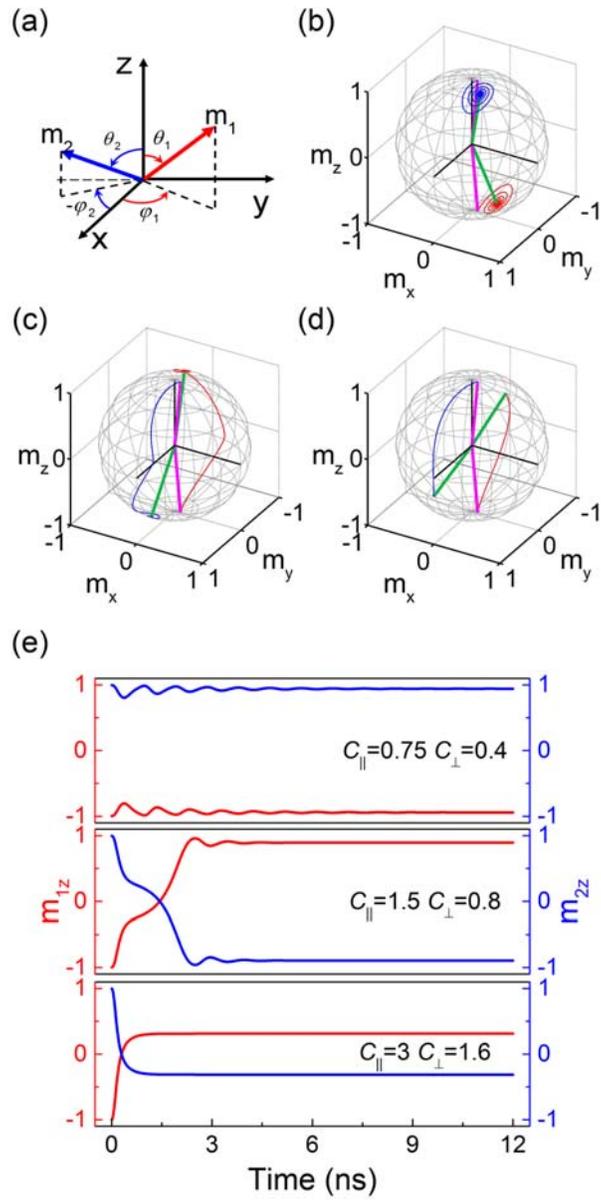

Shi *et al.*, FIG. 4.